\documentclass[12pt,preprint]{aastex}
\usepackage{emulateapj5}
\usepackage{apjfonts}
\usepackage{epsfig}
\usepackage{natbib}
\newcommand{\chisq}{\ensuremath{\chi^2}}

\newcommand{\etal}{et al.}

\newcommand{\delsig}{\ensuremath{\Delta \sigma}}

\newcommand{\dn}{D\ensuremath{_n}(4000)}

\newcommand{\flamb}{erg s$^{-1}$ cm$^{-2}$ \AA$^{-1}$}

\def\gtrsim{\mathrel{\hbox{\rlap{\hbox{\lower4pt\hbox{$\sim$}}}\hbox{\raise2pt\hbox{$>$}}}}}

\newcommand{\fwoiii}{\ensuremath{\mathrm{FWHM}_\mathrm{[O {\tiny III}]}}}
\newcommand{\fwoii}{\ensuremath{\mathrm{FWHM}_\mathrm{[O {\tiny II}]}}}

\newcommand{\halpha}{H\ensuremath{\alpha}}

\newcommand{\hbeta}{H\ensuremath{\beta}}

\newcommand{\hd}{H\ensuremath{\delta_{\rm A}}}

\newcommand{\hst}{\emph{HST}}

\newcommand{\kms}{km~s\ensuremath{^{-1}}}

\newcommand{\lf}{\ensuremath{L_{\rm{5100 \AA}}}}

\newcommand{\lum}{erg s$^{-1}$}

\newcommand{\lledd}{\ensuremath{L_{\mathrm{bol}}/L{\mathrm{_{Edd}}}}}

\newcommand{\loiii}{\ensuremath{L_{\mathrm{[O {\tiny III}]}}}}

\newcommand{\mbh}{\ensuremath{M_\mathrm{BH}}}

\newcommand{\mgb}{\ion{Mg}{1}$b$}

\newcommand{\msigma}{\ensuremath{M_{\mathrm{BH}}-\sigmastar}}
\newcommand{\msun}{\ensuremath{M_{\odot}}}

\newcommand{\oii}{[\ion{O}{2}]}
\newcommand{\oiii}{[\ion{O}{3}]}

\newcommand{\sers}{S{\'e}rsic}

\newcommand{\sigmastar}{\ensuremath{\sigma_{\ast}}}
\newcommand{\sigmagas}{\ensuremath{\sigma_{\rm g}}}

\newcommand{\whz}{W~Hz$^{-1}$}

\def\lax{{$\mathrel{\hbox{\rlap{\hbox{\lower4pt\hbox{$\sim$}}}\hbox{$<$}}}$}}
\def\gax{{$\mathrel{\hbox{\rlap{\hbox{\lower4pt\hbox{$\sim$}}}\hbox{$>$}}}$}}

\slugcomment{Draft version 11; June 6, 2009;
to be submitted to {\it The Astrophysical Journal}.}
\shorttitle{{\it Obscured Active Galaxies}}
\shortauthors{GREENE ET AL.}

\begin{document}

\title{The Growth of Black Holes: Insights From Obscured Active Galaxies}

\author{Jenny E. Greene}
\affil{Department of Astrophysical Sciences, Princeton University, 
Princeton, NJ 08544; Hubble, Princeton-Carnegie Fellow}

\author{Nadia L. Zakamska}
\affil{Institute for Advanced Study, Einstein Dr., Princeton, NJ 08540;
Spitzer Fellow; John N. Bahcall Fellow}

\author{Xin Liu}
\affil{Department of Astrophysical Sciences, Princeton University, 
Princeton, NJ 08544}

\author{Aaron J. Barth}
\affil{Department of Physics and Astronomy, 4129 Frederick Reines Hall, 
University of California, Irvine, CA 92697-4575}

\author{Luis C. Ho}
\affil{The Observatories of the Carnegie Institution of Washington, 
813 Santa Barbara Street, Pasadena, CA 91101}

\begin{abstract}

Obscured or narrow-line active galaxies offer an unobstructed view of
the quasar environment in the presence of a luminous and vigorously
accreting black hole.  We exploit the large new sample of optically
selected luminous narrow-line active galaxies from the Sloan Digital
Sky Survey at redshifts $0.1 < z < 0.45$, in conjunction with
follow-up observations with the Low Dispersion Survey Spectrograph
(LDSS3) at Magellan, to study the distributions of black hole mass and
host galaxy properties in these extreme objects.  We find a
narrow range in black hole mass ($\langle$log \mbh/\msun$\rangle = 8.0
\pm 0.7$) and Eddington ratio ($\langle$log \lledd$\rangle = -0.7 \pm
0.7$) for the sample as a whole, surprisingly similar to comparable
broad-line systems.  In contrast, we infer a wide range in star
formation properties and host morphologies for the sample, from
disk-dominated to elliptical galaxies.  Nearly one-quarter have highly
disturbed morphologies indicative of ongoing mergers.  Unlike the
black holes, which are apparently experiencing significant growth, the
galaxies appear to have formed the bulk of their stars at a previous
epoch.  On the other hand, it is clear from the lack of correlation
between gaseous and stellar velocity dispersions in these systems that
the host galaxy interstellar medium is far from being in virial
equilibrium with the stars.  While our findings cast strong doubt on
the reliability of substituting gas for stellar dispersions in high
luminosity active galaxies, they do provide direct evidence that
luminous accreting black holes influence their surroundings on a
galaxy-wide scale.

\end{abstract}


\section{Introduction}

Over recent years the cosmological status of accreting supermassive
black holes (BHs) has been elevated from an extreme rarity to a
ubiquitous phase in the life cycle of all massive galaxies
\citep[e.g.,][]{kormendyrichstone1995}.  There appears to be direct
evidence for a connection between BH and galaxy growth in the tight
scaling relations between BH mass and bulge properties \citep[e.g.,
the \msigma\ relation;][]{tremaineetal2002}.  Nonetheless, all of the
classic conundrums regarding BH remain.  There is not yet a reliable
measurement of the length of typical BH growth episodes
\citep{martini2004}, or the typical mode of BH growth
\citep[e.g.,][]{kollmeieretal2006,shenetal2008,gavignaudetal2008}.
Furthermore, while it has been argued both on observational
\citep[e.g.,][]{sandersmirabel1996,canalizostockton2000,gonzalezdelgadoetal2001}
and theoretical \citep{mihoshernquist1994,goodman2003} grounds that
star formation and accretion activity are temporally coincident,
quantitative stellar population constraints for large samples of
accreting BHs at high luminosities are notoriously difficult to obtain
\citep[e.g.,][]{borosonoke1984,
bahcalletal1997,canalizostockton2001,letaweetal2007,jahnkeetal2007}.
The accretion power from a luminous active galactic nucleus (AGN) may
easily outshine the surrounding host galaxy by a factor of 100, which
makes studies of AGN host galaxy stellar populations and morphologies
a considerable challenge.

Occasionally nature is particularly kind and puts obscuring material
between us and the accretion disk.  In these cases, we know that an
accreting BH is present due to the ratios of strong emission lines
that imply a very hard ionizing spectrum
\citep[e.g.,][]{baldwinetal1981}, and BHs selected this way are known
as narrow-line or Type 2 AGNs.  These objects are an ideal tool with
which to study AGN host galaxies, since the galaxy to AGN contrast
ratio is maximized.  Much of what we know about the stellar
populations of AGN host galaxies is derived from narrow-line
objects \citep[e.g.,][]{heckmanetal1997,gonzalezdelgadoetal2001,
  cidfernandesetal2004}.  However, until recently there were not large
samples of homogeneously selected, {\it luminous} narrow-line AGNs.
The situation has been remedied by the large spectroscopic database
provided by the Sloan Digital Sky Survey \citep[SDSS;][]{yorketal2000}.
\citet{zakamskaetal2003} used the strong and ubiquitous \oiii$~\lambda
5007$ line emission as a proxy for bolometric luminosity in order to
select a sample of narrow-line AGNs as luminous as any known broad-line
active galaxies with $z<1$.  More recently \citet{reyesetal2008}
presented a sample of nearly 1000 narrow-line AGNs with $z <
0.8$ and bolometric luminosities as high as $10^{47}$~erg~s$^{-1}$ or more.

We have initiated a program to obtain deeper follow-up long-slit
spectroscopy of the Reyes \etal\ sample at $z < 0.45$.  Our primary
goal is to study the evolutionary state of the host galaxies of
vigorously accreting BHs.  Both stellar velocity dispersions
(\sigmastar) and stellar population age can be measured from the
stellar continua.  From \sigmastar\ we derive not only a mass-scale
for the galaxy, but also an estimate of the BH mass and in turn
Eddington fractions for the sample, assuming that
the \msigma\ relation holds \citep[which may not be the case for active
galaxies at these redshifts; e.g.,][]{wooetal2006,kimetal2008b}.
By comparing the structural properties of the hosts to those of
inactive galaxies at a similar mass, combined with our measurements of
stellar ages, we can hope to determine whether the host galaxies are
evolving coevally with their BHs.  Of course, as we discuss in 
\S 7, it is not yet clear whether broad- and narrow-line AGNs are 
truly identical populations seen at different orientations 
\citep[as suggested by unification;][]{antonucci1993}, or whether they 
represent an evolutionary sequence \citep[as suggested by the differences 
in star formation properties, among other things, e.g.,][]{kimetal2006,
hopkinsetal2006,zakamskaetal2008}.

Throughout we assume the following cosmological parameters to calculate
distances: $H_0 = 100~h = 70$~\kms~Mpc$^{-1}$, $\Omega_{\rm m} = 0.30$,
and $\Omega_{\Lambda} = 0.70$.

\section{The Sample, Observations, and Data Reduction}

Our goal was to select nearby targets, for which we could hope to
resolve the emission-line structure, that are nevertheless at the
upper end of the luminosity function.  We thus limited our attention
to targets with $z<0.45$, such that \oiii\ is accessible in the
observing windows (see below) and with $\loiii \geq
10^{42}$~erg~s$^{-1}$.  While arbitrary, our luminosity cut is chosen
to match the classic definition of a quasar.  Altogether, we obtained
spectroscopic observations for 15 targets.  Our sample spans, by
construction, a much narrower range in redshift (and thus luminosity)
than the parent sample.  

All of the data discussed in this paper were collected over two
observing runs using the Low Dispersion Survey Spectrograph (LDSS3) at
the Clay-Magellan telescope on Las Campanas.  LDSS3 is a wide-field
(8.3\arcmin\ diameter) imaging spectrograph, which was upgraded from
the original LDSS2 \citep{allington-smithetal1994} to have increased
sensitivity in the red. For each target we obtained 2--5 min
acquisition images in $g$ and $r$, and sometimes $i$.  We then spent
at least 1 hr per target at the primary slit position, as well as
additional time as noted in Table 1, using a 1\arcsec$\times$4\arcmin\
slit.  Slit position angles were chosen by manual inspection of the
SDSS color images, with an eye to investigate color gradients or
low-surface brightness material (Table 1).  The seeing ranged from
0\farcs6 to 1\farcs5 over the two runs, but was typically 1\arcsec;
with a plate scale of 0\farcs189 the spectra are well sampled in the
spatial direction.  Two spectroscopic settings were required in order
to ensure that \oiii\ was observed over the full
redshift range of our sample.  The primary setting was the reddest
position of the VPH-Blue grism, which covers $4300-7050$ \AA.  For the
highest redshift targets, we needed a redder setting, and used the
bluest setting of the VPH-Red grism, which covers $5800-9400$ \AA. Our
velocity resolution was $\sigma_{\rm inst} \approx 67$~\kms\ for both
settings.  Each night, in addition to the science targets, we observed
two to four spectro-photometric standards covering a range of airmass.
Finally, over the course of the runs we also observed a large library
of velocity template stars, consisting mostly of G--M giants.

The data reduction followed standard 
procedures\footnote{http://iraf.net/irafdocs/spect/}, with the exception of
pattern-noise removal as discussed below.  We first performed cosmic-ray
\hskip 0.2in
\psfig{file=tableobsv3.epsi,width=0.45\textwidth,keepaspectratio=true,angle=0}
\vskip 4mm
\noindent
removal on each individual frame using the spectroscopic version of
LACosmic \citep{vandokkum2001}.  We performed bias-subtraction,
flat-fielding, wavelength calibration, pattern noise removal (when
necessary; see below) and rectification using the Carnegie
Observatories reduction package
COSMOS\footnote{http://www.ociw.edu/Code/cosmos}.  The rectification
works extremely well; in the rectified images the trace centroid
typically migrates by less than two spatial pixels across the full
spectrum, with the maximum drift being five pixels.  COSMOS optionally
performs sky subtraction prior to rectification \citep{kelson2003},
but we have found that for the purposes of stellar velocity dispersion
measurements we achieve significantly better sky subtraction using
{\it apall} in IRAF on the rectified, but not sky-subtracted, frames.
We have used several apertures for spatial extractions, but for
most of this paper we focus on a 2\farcs25 spatial extraction, which
is well-matched to the point-spread function (PSF) for the majority of
our observations. Generally we did not use optimal extraction
\citep{horne1986} in order to avoid emission-line
clipping.

Once the spectra were extracted, we performed flux calibration, minor
shifts to the wavelength calibration and telluric absorption
corrections using IDL routines, following methods described by
\citet{mathesonetal2008}. These routines provide considerable
flexibility and control in the selection of bandpasses for
spectrophotometric calibration used to perform a spline fit to
the continuum.  Small corrections are made, if necessary, to the
wavelength solutions using a cross-correlation with sky lines,
and a heliocentric correction is applied.  Our spectrophotometric
standards are typically white dwarfs, and thus the smooth continuum is
ideally suited to telluric absorption removal as well
\citep[e.g.,][]{wadehorne1988,mathesonetal2000}.  Below we show that
our flux calibration agrees reasonably well with the SDSS by comparing
flux measurements of the \oiii\ line.

The final spectra have a considerably higher signal-to-noise ratio
(S/N) than the SDSS spectra.  For $z < 0.25$, the median S/N at
$5080-5400$~\AA\ is 150 per pixel, which is typically 4 times
higher than the SDSS, while for $z > 0.25$, the median S/N over the
same region is 11 per pixel and 8 times higher than the SDSS
spectra.

\begin{figure*}
\vbox{ 
\vskip -10mm
\hskip 0.6in
\psfig{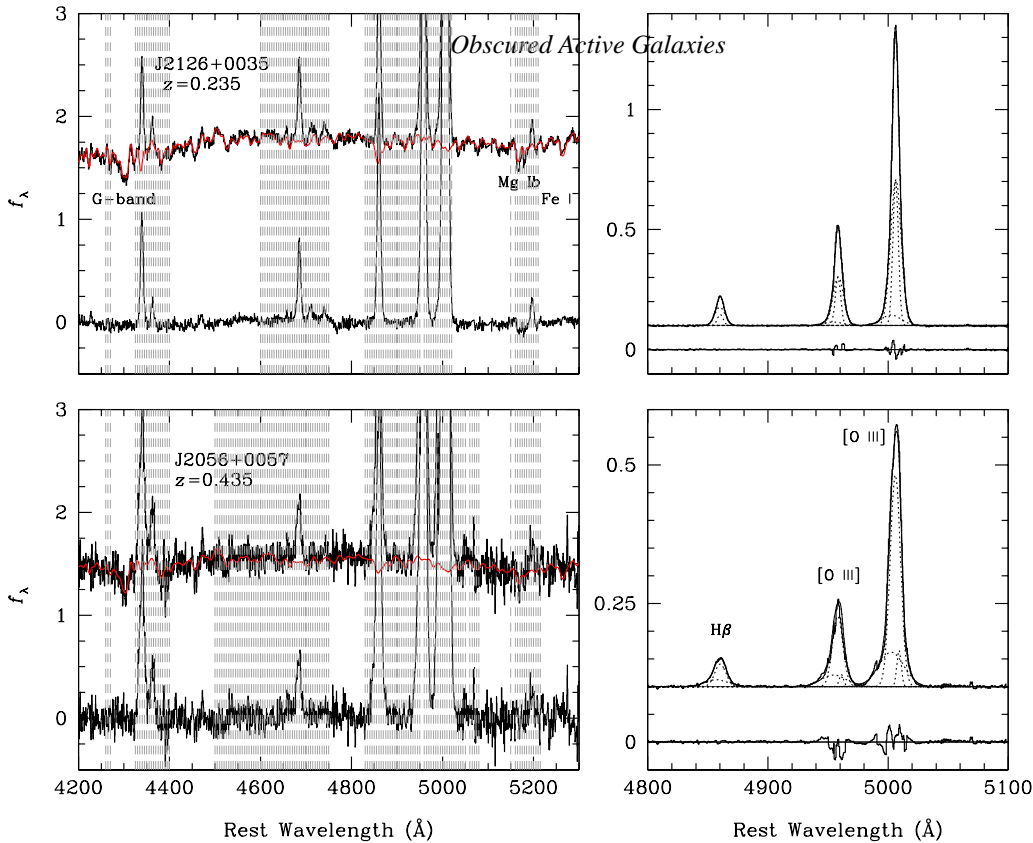}
}
\vskip -0mm
\figcaption[]{
Example fits to two objects in the sample, spanning the range 
of redshift and S/N in the sample.  
{\it Left}: Stellar velocity dispersion fits using the best weighted
fit of an F2, G5, K1, and K4 star.  We plot LDSS3 spectra ({\it thick
histogram}), best-fit model ({\it red solid}), and residuals ({\it
thin histogram}). Data and models are offset for clarity, while
regions excluded from the fit are demarcated with grey hatched lines.
These fits provide both a measurements of \sigmastar\ and a 
continuum-subtracted spectrum.  Flux density is normalized to the continuum.
{\it Right}: Fits to the \hbeta\ and [O {\tiny III}] emission lines using 
multi-component Gaussians; see text for details. Total model 
({\it thin solid}) and model components ({\it dotted}) are plotted over 
the continuum-subtracted data ({\it thick histogram}).  Residuals are 
offset for clarity and flux density is in units of $10^{-15}$~\flamb.
\label{fits}}
\end{figure*}

LDSS3, particularly during the March 2007 observation period, had
intermittent but substantial pattern noise.  Pattern noise appears as
periodic striping in the data, with a typical scale of a few pixels
and an amplitude (in these data) of 50--100 counts.  The amplitude and
phase vary in time and so standard calibrations do not remove it but
there is a COSMOS task that Fourier transforms the data and removes
the high-frequency noise.  Our concern is that the frequency of the
pattern noise is comparable in scale to that of the expected stellar velocity
dispersions, potentially leading to biases in our final measurements.
In order to test this possibility, we have performed three independent
pattern-noise removal runs for the first night, which has particularly
bad pattern noise.  We have measured the dispersions independently for
each of these spectra.  The rms in the measured \sigmastar\ values for
the four objects range from 3 to 20 km~s$^{-1}$, and, most crucially,
we see no systematic differences in the measured dispersion for
different versions of the pattern-noise removal.  We conservatively
add an additional 20~km~s$^{-1}$ uncertainty in quadrature to the
measured dispersions that suffer from pattern noise.

\section{Stellar Velocity Dispersions}

One of the primary uncertainties in interpreting observations of
narrow-line quasars comes from their unknown BH masses.  For the
median \loiii$=2 \times 10^{42}$~erg~s$^{-1}$ of the Magellan sample,
we estimate that the bolometric luminosity is $L_{\rm bol} \approx 2
\times 10^{45}$ erg s$^{-1}$, which suggests BH masses of $>
10^7$~\msun\ if they emit at or below their Eddington limits [where we
take the Eddington luminosity to be $1.26\times10^{38}$~(\mbh/\msun) \lum].
However, given the scatter between \oiii\ and bolometric luminosity,
the range of bolometric luminosities in the sample, and their
(presumed) range of intrinsic Eddington ratios, it would be nice to
have an independent estimate of the BH masses.  At the same time, it
would be very useful to have a mass scale for the host galaxies.
Although it is indirect, if we assume that the \msigma\ relation
holds, then we can use measurements of bulge velocity dispersions both
to provide an estimate of the BH masses in the sample and as a proxy
for the mass of the host galaxy \citep[e.g.,][]{heckmanetal2004}.
Here we describe the \sigmastar\ measurements and uncertainties; the
BH masses are presented in \S 5.

There are a variety of commonly used techniques for measuring stellar
velocity dispersions in galaxies.  For computational expediency, most
of them rely on Fourier techniques, including cross-correlation
\citep{tonrydavis1979}, the Fourier quotient method
\citep{simkin1974,sargentetal1977}, and the Fourier correlation
quotient \citep{bender1990}.  It is also possible to directly compare
broadened templates with data in pixel space
\citep[e.g.,~][]{burbidgeetal1961}.  Direct-pixel fitting methods,
while relatively computationally expensive, provide many benefits over
Fourier techniques \citep[e.g.,~][]{rixwhite1992, vandermarel1994,
  kelsonetal2000, barthetal2002, bernardietal2003b}, including the
ability both to fit over restricted regions and to mask complicating
emission or absorption features.  As described in detail in
\citet{greeneho2006sig}, even faint emission lines can lead to
substantial biases in derived \sigmastar\ measurements if not properly
accounted for.

Our galaxy models are built from a linear combination of stellar
spectra.  The program object is de-redshifted using the SDSS redshift, 
but inevitably small velocity shifts are present between the templates 
\hskip -0.2in
\psfig{file=tablegalv5.epsi,width=0.5\textwidth,keepaspectratio=true,angle=0}
\vskip 4mm
\noindent
and the target.  Thus, each template is shifted to zero velocity, diluted by a
constant or power-law component $C(\lambda)$, convolved with a
Gaussian $G(\lambda)$, and multiplied by a polynomial $P(\lambda)$:
\begin{equation}
M(\lambda) = P(\lambda) \{ [ T(\lambda) \otimes G(\lambda)] + C(\lambda) \}
\end{equation}
The nonlinear Levenberg-Marquardt algorithm, as implemented by {\tt mpfit}
in IDL \citep{markwardt2009}, is used to minimize \chisq\ in pixel
space between the galaxy and the model.  Our code is based on that
presented in \citet{greeneho2006sig}, except that the model spectrum
is actually a linear combination of stars (typically four in this
work) with their relative amplitudes as additional free parameters.
This version of the code was developed to model the high-quality
nuclear spectra from the Palomar spectroscopic survey of nearby
galaxies \citep{hoetal1995} and is presented and extensively tested in
\citet{hoetal2009}.  Unless otherwise noted we have used the following
four stars as our template set: HD 26574 (F2~III), HD 107950 (G5~III),
HD 18322 (K1~III), and HD 131507 (K4~III).  Example fits are shown in
Figure 1.  While the primary purpose of the multi-template fitting is
to achieve reliable fits, we also extract crude stellar population
information in \S 7.

In order to use the full spectral range of the observations, we cannot
rely only on our own template stars, whose wavelength coverage does
not go blueward of $\sim$4300 \AA.  Rather, we utilize a large library
of high-S/N, high-spectral resolution ($\sim 26$~\kms) stellar
templates from \citet{valdesetal2004}.  It is best to use template
stars observed with an identical instrumental setup as the targets, so
that the measured broadening can be ascribed solely to internal
kinematics in the galaxy.  However, the Valdes templates not only
allow us to include spectral features such as the G-band $\lambda
4304$~\AA\ to the highest redshifts, but also provide a wide range of
spectral types to help minimize the impact of template mismatch.  We
thus use a simple bootstrapping procedure to remove the impact of
differing instrumental resolutions on the measured dispersions.  First
of all, we measure the intrinsic broadening of the

\psfig{file=cfsigma.epsi,width=0.44\textwidth,keepaspectratio=true,angle=0}
\vskip -0mm
\figcaption[]{
Comparison of stellar velocity dispersion measurements using our 
internal velocity template stars ($\sigma_{\rm Our}$) and the 
Valdes template stars ($\sigma_{\rm Valdes}$) for the low-redshift 
targets.  The template stars, while not identical, are matched exactly 
in spectral type to remove ambiguity caused by template mismatch. 
The measured dispersions from the Valdes templates are 
corrected for resolution assuming that the instrumental resolution 
of our LDSS3 measurements is $\sigma$=67~\kms, while the instrumental 
resolution of the Valdes stars is $\sigma$=26~\kms.  As demonstrated 
here, the agreement is excellent, which supports our assumed conversion 
factor.
\label{cfsigma}}
\vskip 5mm
\noindent
LDSS3 template stars using the Valdes library.  While we did not
observe identical stars, we can match the stars in spectral type
nearly exactly.  We expect the relative broadening to be $\sigma_{\rm
  rel} = (\sigma^2_{\rm LDSS3}-\sigma^2_{\rm Valdes})^{0.5} =
(67^2-26^2)^{0.5} = 62$~\kms.  We measure a mean broadening over 10
stars of 60$\pm$7~\kms.  Dividing the sample into the two runs
demonstrates the stability of the spectrograph; we find 61$\pm$8~\kms\
and 59$\pm$7~\kms, respectively.  Here and throughout, when we quote
errors on mean quantities, we are quoting the standard deviation of
the distribution rather than the error on the mean.

We then confirm that our conversion works properly by measuring the
velocity dispersions of our targets with the Valdes templates and
comparing the values to those measured from our internal templates,
again being careful to match the spectral types of the template stars
and the fitting regions.  The resulting dispersions are shown in
Figure 2, where the reported Valdes dispersions have been reduced by
$60$~\kms\ in quadrature.  It appears that our conversion is quite
robust.  From now on we report results based entirely on the Valdes
stars with the above correction, conservatively adding 7 \kms\ in
quadrature to our error budget. The resulting \sigmastar\ measurements
are shown in Table 2.

In the interest of uniformity, in all cases we use the spectral region
4100--5400 \AA, which is accessible for every observation.  
We follow \citet{greeneho2006sig} and mask out the \mgb~$\lambda
\lambda 5167, 5172, 5183$ triplet.  
While the \mgb\ triplet is one of
the highest equivalent width (EW) features in this spectral region,
the well-known correlation between the [Mg/Fe] ratio and \sigmastar\
\citep[e.g.,][]{oconnell1976,kuntschneretal2001} leads to a systematic
overestimate of \sigmastar\ based on \mgb\ \citep{barthetal2002}.  

\psfig{file=ddelsig_k1.epsi,width=0.45\textwidth,keepaspectratio=true,angle=0}
\vskip -0mm
\figcaption[]{
Input-output simulations to estimate uncertainties in
\sigmastar\ resulting from the finite S/N and continuum dilution.
\delsig$\equiv \langle \sigma_{\rm out} - \sigma_{\rm in} \rangle /
\sigma_{\rm in}$.  We investigate S/N=10 (solid black, circles),
S/N=15 (blue short-dashed, triangles), and S/N=30 (red long-dashed,
squares).  In all cases, two-thirds of the continuum is a nonthermal
component modeled as a constant.  Each setting is run with 10
randomly generated error arrays.  We see that there is no net offset
even at these low S/N ratios and high dilution values, but
uncertainties range from $\sim 10\%$ to 20\%.
\label{cfsigma}}
\vskip 5mm

\subsection{SDSS Spectra}

In order to increase our sample size, we have also considered all of
the targets in the \citet{reyesetal2008} sample.  Since most of the
spectra do not have the required S/N for reliable velocity dispersion
measurement, we have limited our attention to objects with a median
S/N of $>$15 per pixel between 5080 and 5400 \AA, and an EW in the Ca K line
greater than 3 \AA.  There are 547 objects with $L_{\rm [O \tiny III]} >
10^{42}$~erg~s$^{-1}$; of those, 111 make our cut.  The SDSS velocity 
dispersions are measured in an identical fashion as the fits described 
above, using the same mix of Valdes template stars.  In this case, we 
assume an instrumental resolution of 71 \kms\ for the SDSS spectra 
\citep[see discussion in][]{greeneho2006sig}.

The SDSS-derived \loiii\ values, our measured velocity dispersions,
and \oiii\ line widths for this additional sample are all presented in
Table 3.  Two objects are unresolved (\sigmastar$<71$~km~s$^{-1}$).
There are only two objects from the Magellan sample for which we can
measure reliable dispersions from the SDSS spectra, and the resulting
dispersions agree with ours within the quoted uncertainties.

\subsection{Uncertainties}

Our quoted uncertainties are a quadrature sum of the formal errors
from the fit and the 7~\kms\ uncertainty incurred in converting to the
Valdes system.  Our typical uncertainties thus derived range from
$10\%$ for the lower-redshift targets to $40\%$ for the most distant
and luminous targets.  

As the S/N in the spectrum decreases, our ability to recover the true
dispersion decreases.  Also, an increasing contribution to the
continuum from sources apart from old stars lowers their EW and thus
the S/N.  To a small degree the relative incurred error also depends
on the true dispersion.  

\hskip -0.1in
\psfig{file=tableshv2.epsi,width=0.45\textwidth,keepaspectratio=true,angle=0}
\vskip 4mm
\noindent
We run a suite of simulations designed to
quantify each of these effects by adding Gaussian random noise and a
constant continuum of various levels to one of the template stars that
has been broadened to various widths.  We create ten mock spectra for
each S/N value, and then measure \sigmastar\ for each mock spectrum.
This simple input-output experiment shows that our error bars are
reasonable.  We expect $10\%-20\%$ errors for the targets with $10 <
S/N < 30$ and $67\%$ continuum dilution (Fig. 3).

Systematic uncertainties may be introduced in the gas and stellar
dispersions due to our use of fixed angular rather than physical
apertures.  In the case of the stars, we are typically extracting the
spectra at $\sim 0.5~r_{\rm e}$.  Since the velocity dispersion
profiles of bulge-dominated galaxies are known to be flat
\citep[e.g.,][]{jorgensenetal1995}, in general aperture bias should be
small compared to the measurement uncertainties.  The only potential
exceptions are the disk-dominated galaxies (J1106, J1222, and J1253)
for which disk contamination may be significant.  In the case of the
gas, line widths can change dramatically as a function of aperture
(typically getting narrower at large radius).  Unfortunately, the
dependence of gas dispersion on radius varies substantially from
object to object \citep[see, e.g.,~][]{riceetal2006,walshetal2008} and
there is no physically motivated aperture to choose.  In future work
we will explore the full variation in gas velocity dispersion and
luminosity as a function of radius.

\subsection{Emission-line Fits to Nuclear Spectra}

A convenient additional product of our velocity dispersion fits is a
continuum-subtracted emission-line spectrum for each target.  Since we
are interested in the \oii\ line when it falls in the bandpass, we
extend the model out to the bluest available pixels although we do not
use those wavelengths in the fitting.  Examples of the
continuum-subtracted spectra are shown in Figure 1, and we note that
all spectra have been extracted with a 2\farcs25 aperture.  We fit the
\hbeta, \oiii, and \oii\ lines with multi-component fits.  Briefly,
the \oiii~$\lambda\lambda 4959, 5007$ lines are modeled with identical
sets of Gaussian components, where the centroid shifts are fixed to
the laboratory value.  The \hbeta\ line is fit with a combination of
two Gaussians (but see \S7 for an alternate fit).  Our prescription
follows that described in detail in Greene \& Ho (2005a), except that
unlike that work in general we require more than two components to
adequately model the lines.  Furthermore, we do not place any
restrictions on the relative widths or strengths of the lines.  The
\oii~$\lambda \lambda 3726, 3729$ lines are modeled with identical
sets Gaussian components (usually two), and the line ratios in the
doublet are fixed to one.  This procedure is necessary to decouple
true velocity dispersion from the small velocity separation of the two
components of the doublet.

The line widths and strengths are shown in Table 4.  The
\oiii\ line is known to have an asymmetric profile, often with a
prominent blue wing \citep[e.g.,][]{heckmanetal1981}.
\citet{greeneho2005o3} find that using a two-component fit and
removing the blue wing results in a line width that better tracks
\sigmastar.  However, we find that in many cases a two-component model
is not an adequate description of these lines.  As shown by many
authors \citep[e.g.,][]{greeneho2005o3,barthetal2008a,ho2009o3}, the
FWHM of the line is relatively free of bias from the velocity structure 
at the base of the line.
Therefore, we simply take \sigmagas=FWHM/2.35, as measured from our
best-fit emission-line model.

Although in the bulk of this paper we focus on the extractions with a
2\farcs25 aperture, we have made a variety of other extractions, with
widths of 3\farcs78, 5\farcs67, and 7\farcs56.  The last has an
effective area comparable to the SDSS fiber (7 arcsec$^2$).  In Figure
4 we compare the \loiii\ measurements for our objects extracted
(typically) within a 7\farcs56 aperture with the SDSS measurements.
We find overall reasonable agreement, with $\langle \log L_{\rm [O
  {\tiny III}], SDSS} - \log L_{\rm [O {\tiny III}], our}\rangle =
0.01 \pm 0.2$.  One object is a significant outlier; our measurement
of J2212 is $\sim 6$ times fainter than that derived from the SDSS
spectrum.  This problem persists regardless of the flux calibrator
used.  It is curious, since we do not see a similar problem for any
other target observed on this night.  Since the discrepancy is seen in
both the continuum level and the \loiii\ value, and since J2212 is a
highly disturbed and spatially extended system (\S 4), we suspect that
we were just unlucky in our slit placement and missed some luminous
region of the galaxy.  None of the conclusions presented here are
impacted if we use the SDSS luminosity rather than ours.  If we remove
J2212, then the scatter between the two sets of measurements drops to
0.15 dex.  We note that in our own observations \loiii\ can vary by as
much as $80 \%$ between different aperture positions, and so we are
satisfied with this level of agreement.

\section{Galaxy Structural Measurements}

\subsection{Photometric Zeropoints}

We have also reduced the acquisition images (Table 1) using standard
bias-subtraction and flat-fielding routines within IRAF.  We do not
attempt to remove the pattern noise in the images.  Each frame is tied
to the SDSS photometric system using three to six field stars.  We use
the IRAF package {\it phot} to perform aperture photometry on these 
relatively uncrowded fields, and then calculate a weighted average
zeropoint based on the reported photometric errors in the SDSS
photometry.  The formal uncertainties in the final zeropoints are
$<0.05$ mag.  

\subsection{Photometric Decomposition}

The ground-based imaging reaches a typical surface brightness limit
of $24-25$ mag arcsec$^{-2}$, and it is worth using the
images to derive bulge-to-disk decompositions for the sample.  Our
goal is to isolate the bulge-like component of the host galaxy in
order to compare with the \citet{faberjackson1976} relation of
inactive bulge-dominated galaxies.  The Faber-Jackson relation can
provide a valuable diagnostic as to whether the galaxy has reached the
relaxed state of local elliptical galaxies.

We use the two-dimensional surface brightness fitting program GALFIT
\citep{pengetal2002} to model each galaxy.  Our band of choice is
Sloan $r$, since we have it for the majority of the targets.  However,
for various reasons we are forced in a few cases to model either the
$g$- or $i$-band image.  
\psfig{file=cfsdsso3v2.epsi,width=0.4\textwidth,keepaspectratio=true,angle=0}
\vskip -0mm
\figcaption[]{
Comparison of \loiii\ as measured from the SDSS spectra and our
long-slit data with an aperture (typically) of 7\farcs56.  The
agreement is reasonable, with $\langle L_{\rm [O {\tiny III}], SDSS}
- L_{\rm [O {\tiny III}], our}\rangle = 0.01 \pm 0.2$, and the
scatter drops to 0.15 dex when the outlier (J2212) is excluded.
\label{fits}}
\vskip 5mm
\noindent 
In these cases we convert to $r$ band using the
measured color from the SDSS Petrosian magnitudes.  Throughout,
magnitudes are quoted in the AB system.  

In comparison with broad-line AGNs the modeling is relatively
straightforward in this case since there is no nuclear point source.
As a rule, however, parameter coupling is a major complication
\citep[e.g.,][]{kormendydjorgovski1989}, and thus our operating
principle is to model each galaxy with the minimum number of
components.  We are particularly interested in the bulge components of
these galaxies, and we only introduce a disk component if visible in
the images.  In distinguishing between bulges and bars for
disk-dominated galaxies we use the reduced \chisq\ to determine the
preferred model, but we generally cannot constrain more than two
galaxy components robustly.  For the purposes of this paper we have
not placed rigorous limits on the presence of low surface-brightness
components \citep[cf.][]{greeneho2008}.  
Our basic model is the \citet{sersic1968} function:
\begin{equation}
I(r) = I_e~{\rm exp} \left[ -b_n \left(\frac{r}{r_e}\right)^{1/n}-1 \right],
\end{equation}
\noindent
where $r_e$ is the effective (half-light) radius, $I_e$ is the 
intensity at $r_e$, $n$ is the \sers\ index, and $b_n$ is chosen 
such that 
\begin{equation}
\int_0^{\infty}I(r) 2 \pi r dr = 2 \int_0^{r_e} I(r) 2 \pi r dr.
\end{equation}
We adopt the analytic approximation for $b_n$ from \citet{macarthuretal2004}, 
as adapted from \citet{ciottibertin1999}:  
\begin{equation}
b_n \approx 2n-\frac{1}{3}+\frac{4}{405n}+\frac{46}{25515n^2}+
\frac{131}{1148175n^3}-\frac{2194697}{30690717750n^4}.
\end{equation}
The \sers\ model reduces to an exponential profile for $n=1$ and a
\citet{devaucouleurs1948} profile for $n=4$.  Bars may be modeled as
ellipsoids with very low axial ratios.  We follow \citet[][; see also
\citeauthor{freeman1966}~\citeyear{freeman1966}]{dejong1996} and model
the intensity distribution in the bar as a Gaussian.  In total for a
given \sers\ component, the model parameters include the
two-dimensional centroid, the total magnitude, the \sers\ index, the
effective radius, the position angle, and the ellipticity (all of
which are constants with radius for a given component in this version
of GALFIT).  The sky is modeled as a constant pedestal offset
determined using multiple boxes placed in empty regions of the image,
and Levenberg-Marquardt minimization of \chisq\ in pixel space is used
for parameter estimation.

\begin{figure*}
\psfig{file=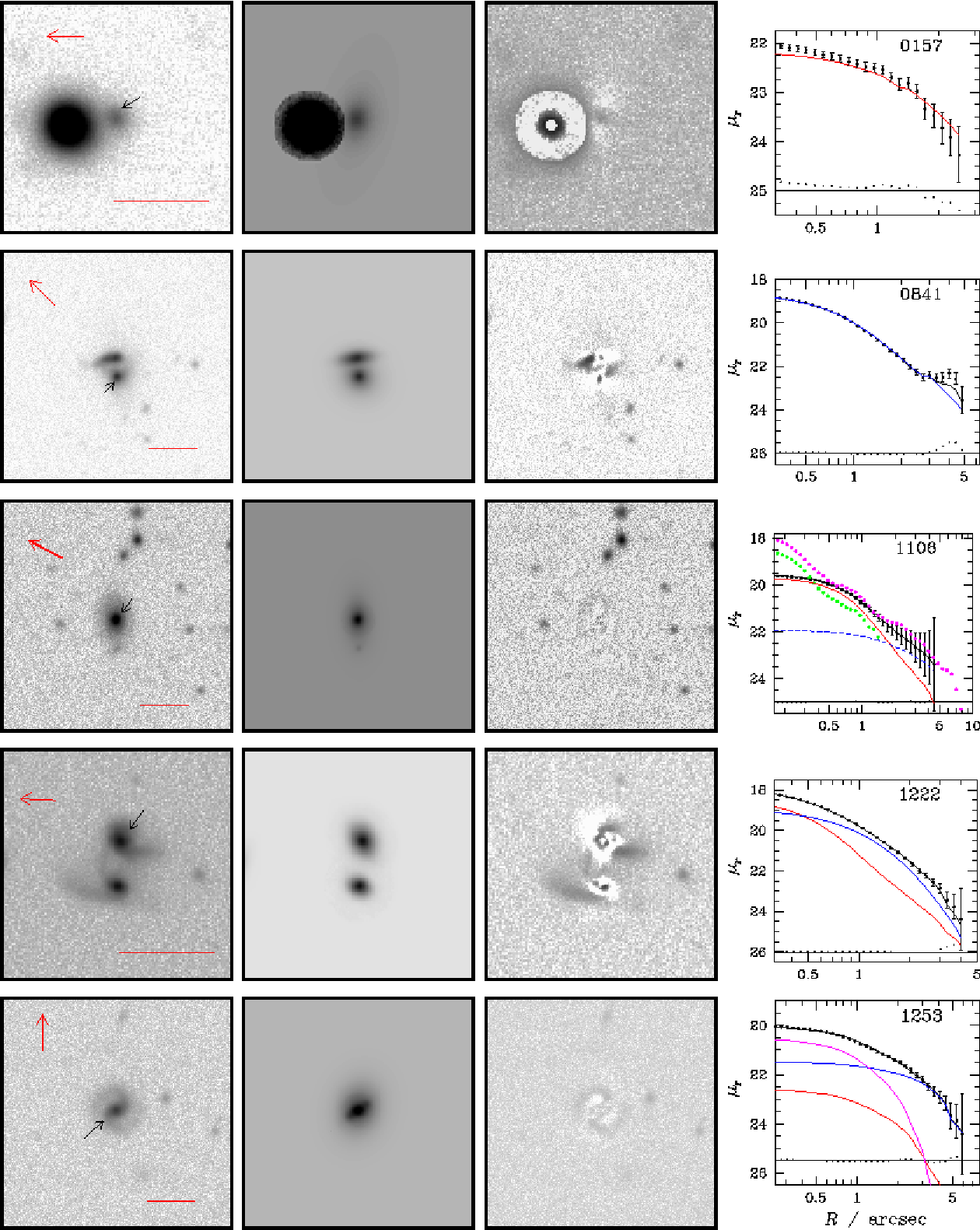,width=0.9\textwidth,keepaspectratio=true,angle=0}
\end{figure*}
\begin{figure*}
\psfig{file=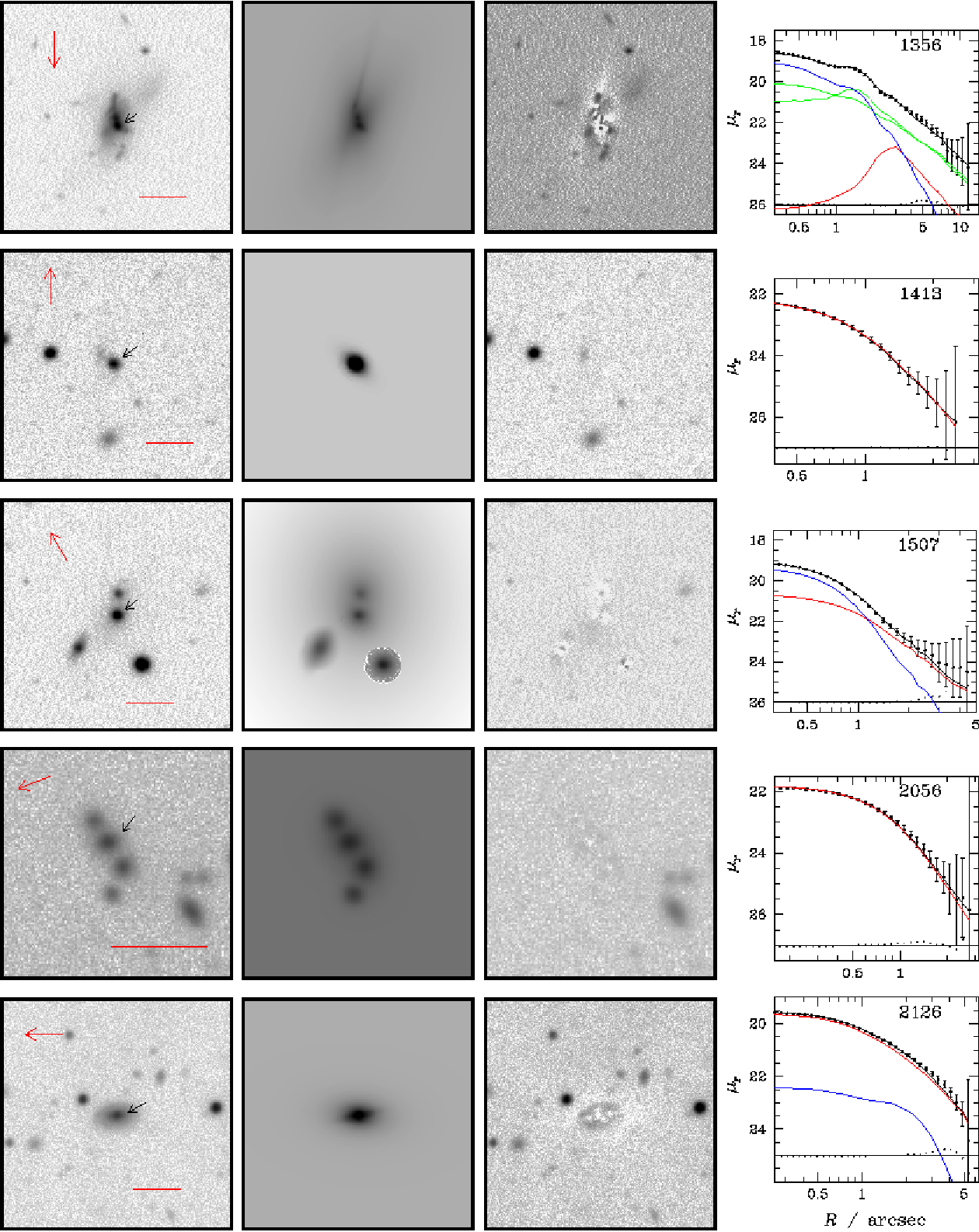,width=0.9\textwidth,keepaspectratio=true,angle=0}
\end{figure*}
\begin{figure*}
\vbox{ 
\vskip -25mm
\hskip 0.in
\psfig{file=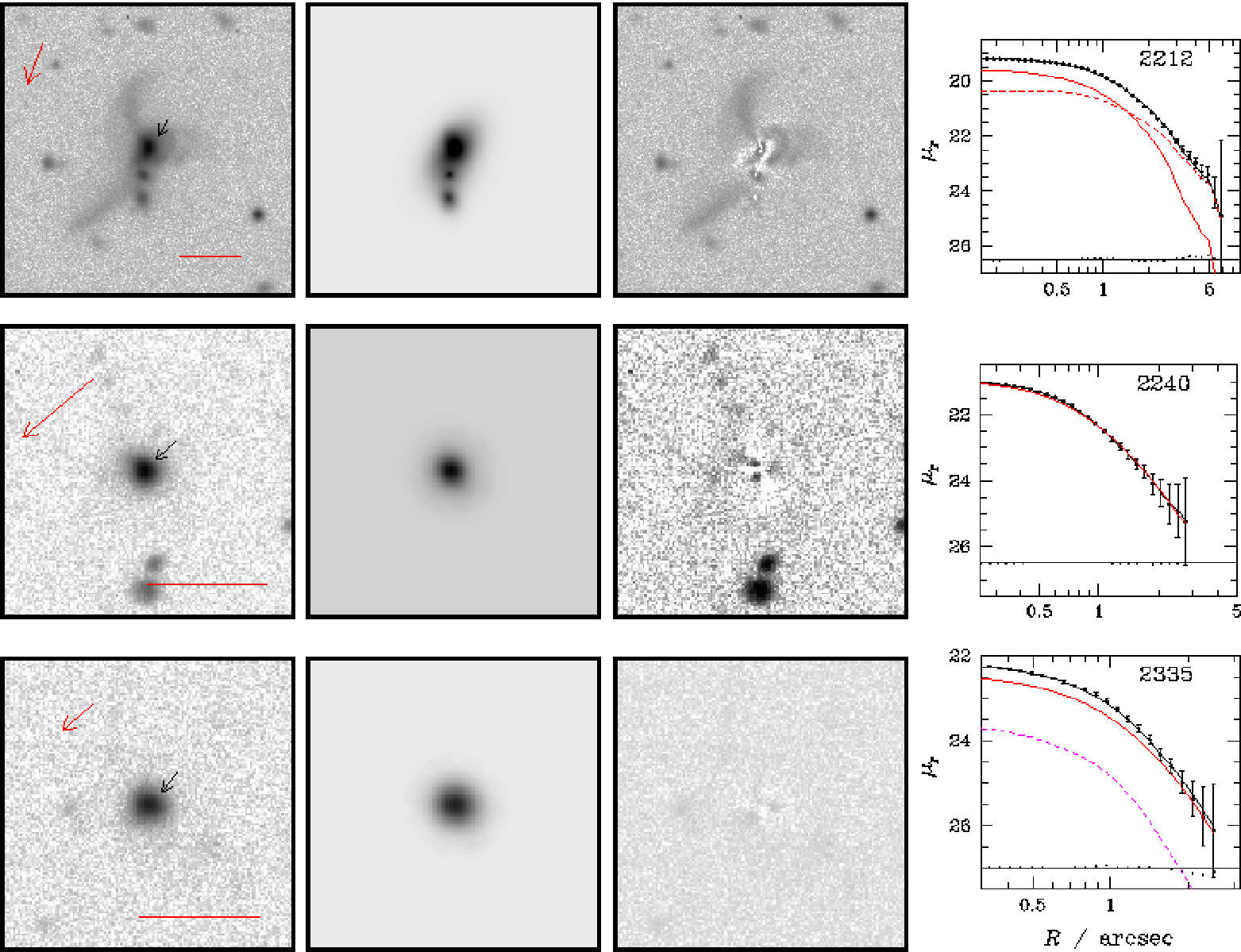,width=0.9\textwidth,keepaspectratio=true,angle=0}
}
\vskip -0mm
\figcaption[]{
Fits are shown for each galaxy, with the
exception of J1124 and J1142, for which the PSF is too poor to use
the images.  Scale bar indicates 10\arcsec, red arrow points North, and
black arrow points to the main target.  We show the LDSS3 image
({\it left}), the best-fit GALFIT model ({\it middle}) and the
two-dimensional residuals ({\it right}), and then the
one-dimensional radial profile.  The points are the data, the red
lines are the bulge components, the blue solid line is the
disk component, the magenta solid line shows a bar component, and magenta
dotted line is the PSF.  The green lines are additional $n=2$
\sers\ components in J1356, which is too disturbed to be well fit by
azimuthally symmetric models such as these.  The one-dimensional
residuals are shown at the bottom.  The magenta dotted line in J2335 
represents an unresolved nuclear component. 
Note that J0157 is in the wings of a very bright star,
making the fitting very uncertain.  The sharp features present 
in the models of J0157 and J1507 are due to the artificial truncation 
of the PSF model.  Finally, in J1106, we compare the ellipse profiles for 
the \hst-ACS FR647M ($\lambda_{\rm c} = 7210$~\AA; {\it magenta points}) 
and F550M ($\lambda_{\rm c} = 5581$~\AA; {\it green points}) with the 
LDSS3 $r$-band image.  Our bulge size may be contaminated by 
a prominent unresolved nuclear source presumably due to scattered light.
\label{fits}}
\end{figure*}

Even without a nuclear point source, proper modeling of the observed
light distribution requires convolution with a PSF.
We follow \citet{ravindranathetal2006} and use {\it Daophot} as
implemented within IRAF to build a PSF model, using no fewer than five
isolated field stars on the same chip as the target galaxy.  The PSF
thus constructed is both noise-free and directly based on the PSF from
the current image.

Two-dimensional modeling provides a few advantages worth noting.  For
one thing, we gain the freedom to assign different position angles and
(if necessary) centroids to different components
\citep[e.g.,][]{wadadekaretal1999}.  Another major benefit of GALFIT
for these galaxies, which often have nearby companions, is that we can
simultaneously model the surrounding galaxies rather than mask them,
which provides a considerably better model of the background.  All
fits are shown in Figure 5.

\subsection{Uncertainties}

Uncertainties in the measured parameters are driven primarily by the
PSF model and the sky level, as well as inadequacies in the assumed
model for the surface brightness profile.  In order to estimate the
former, we adopt a single star from the field as an additional PSF
model.  We further run a pair of models in which the sky level is
increased and decreased by $0.5\%$, which is the typical dispersion in
sky values across the field.  The final error bars represent the
dispersion in each parameter from these four alternate runs.  Our
methodology is described in more detail in \citet{greeneho2008}.

The hardest uncertainties to estimate arise when our assumed model is
incorrect.  Indeed, in J1356 and J2212 (see Table 2 and Fig. 5), the
isophotes are so disturbed due to ongoing merging activity that we do
not achieve a sensible fit from a simple combination of one or two
\sers\ components.  In these two cases, due to the presence of tidal
features and spatially overlapping nuclei, our models are not
particularly meaningful.  Instead, we adopt a nonparametric
\citet{petrosian1976} magnitude for clearly disturbed systems, as
noted in Table 2.  Furthermore, although the direct light from the
active nucleus is extincted by many orders of magnitude, some of it
escapes along unobscured directions, scatters off of the interstellar
matter in the host galaxy and reaches the observer.  Both our
photometric modeling (see below) and spectroscopic fitting (\S 7) may
be affected by this component.  Finally, we do not have usable
Magellan images for J1124 or J1142, and so we adopt SDSS magnitudes in
Table 2.

As a sanity check, we compare our final galaxy luminosities to model
magnitudes from SDSS, as well as to magnitudes from the \hst\ survey
of \citet{zakamskaetal2006} for overlapping galaxies.  The total
magnitudes we measure agree very well with the SDSS model magnitudes,
with an average difference (ours$-$SDSS) of $0.005 \pm 0.15$ mag,
excluding J0157 because the photometry is exceedingly uncertain due to
the proximity of a bright star.  We are thus confident that our
photometry is reasonable, but we adopt an additional 0.15 mag
uncertainty in the bulge luminosities to account for the systematic
differences between the two sets of measurements.  

There are two objects in common between this paper and the \hst\
sample, J1106 and J1413.  Again, the total magnitudes agree reasonably
well, within 0.07 mag in the case of J1106 and 0.4 mag in the case of
J1413.  (Color differences are derived from the LDSS3 spectra, and so
strictly speaking only apply in the inner regions of the galaxy, and
we account only for galactic extinction.)  While we derive similar B/T
ratios for J1106 from the two images, the derived bulge effective
radii are different; we find $r_e = 0.8 \pm 0.4$ kpc, while it is
found to be 3.8 kpc in the nearest \hst\ band (F550M).  The
explanation for this difference can be seen in the radial profiles
derived from the LDSS3 and \hst\ images respectively (Fig. 5).  A
scattered-light component is detected in \hst\ images as a biconical
structure with a maximum extent of 0\farcs9 and appears as a central
brightness peak in the one dimensional brightness profile.  This
component is not resolved with the LDSS3 images but may bias our
derived bulge size to low values.  At the same time, the strong dust
lanes may mimic the transition between bulge and disk in the \hst\
analysis, causing the \hst\ bulge measurement to be too high.  This
one example emphasizes that we are potentially biased in our bulge
sizes and luminosities due to scattered light.  We do not detect the
scattered light signal strongly in the spectrum, presumably because of
internal reddening in the host galaxy.

\subsection{Galaxy Morphology}

In addition to reliable galaxy luminosities, we also have derived
bulge-to-total ratios (B/T; Table 2) and general morphological
information for this sample.  There are four disk galaxies (J1106,
J1222, J1253, and J2126) while the rest consist of a single bulge-like
component, or in the case of J1356 and J2212, train-wrecks.  Based on
SDSS imaging, J1124 appears to be extended, but not obviously disky,
while J1142 is compact.  It is interesting to note, furthermore, that
a large fraction of the sample have companions (eight out of 15, most
confirmed with the long-slit spectroscopy).  Four objects are
significantly tidally disturbed, and thus obviously currently
interacting (J0841, J1356, J1222, and J2212).  There is no correlation
between AGN luminosity or stellar velocity dispersion and galaxy
morphology.  Even if we include the \hst\ observations of
\citet{zakamskaetal2006}, the median \loiii\ for objects with and
without a disk component are identical to within 0.1 dex.  On the
other hand, the sample is quite small.  \citet{zakamskaetal2008} find
a hint that \loiii\ and morphology are correlated, in the sense that
B/T decreases with AGN luminosity, but for objects extending to
somewhat lower luminosities.

\subsection{$k$-corrections and Stellar Populations}

Our ultimate goal is to compare the structural properties of the
narrow-line quasars to those of inactive galaxies, which involves
translating our observations to match the exact observing procedures
and evolutionary state of the local samples.  Since our galaxies span
a redshift range of $0.15 < z < 0.45$, we must assume a model of the
intrinsic spectral shape in order to calculate an effective rest-frame
luminosity.  We use the publicly available program {\it k-correct}
based on the SDSS photometry \citep{blantonroweis2007}.  It would be
preferable to use the bulge component alone to derive the
$k$-corrections, but given our heterogeneous coverage in different
filters for different subsets of objects we use the SDSS photometry in
the interest of uniformity.

Additionally, we must estimate the contamination from emission lines.
These objects can have high EWs in \oiii, meaning that the
line emission contributes significantly to the broad-band luminosity.
From $ 0.1 < z < 0.4$, \oiii\ contaminates the $r$-band magnitude,
while at higher redshift it falls in the $i$ band.  We calculate the
total (extended) \oiii\ luminosity from our widest slit extraction,
and convert to an AB magnitude using {\it synphot} in IRAF.  The
corrections thus derived, as well as the $k$-corrections for the
corrected colors, are shown in Table 2.

Along the same lines, scattered light from the nucleus may contribute
a sizable fraction of the luminosity in the blue.  We have tried to
disentangle contributions from both star formation and scattered light
to the total luminosity (\S 7).  Rerunning {\it
$k$-correct} with the blue continuum removed leads to only $0.02-0.05$
mag of change in the final $k$-corrections.  Finally, of course,
stellar populations fade as they age, and the magnitude of that fading
is dependent on the detailed star formation history of the galaxies in
question.  We will explore this last issue in \S 7.

\section{Black Holes Masses and Eddington Ratios}

\begin{figure*}
\vbox{ 
\vskip -2mm
\hskip 0.4in
\psfig{file=mass_lumall_xinv2.epsi,width=0.4\textwidth,keepaspectratio=true,angle=-90}
}
\vskip -0mm
\figcaption[]{
({\it a}) Sample of narrow-line quasars from \citet{reyesetal2008}
above a luminosity threshold of \loiii$=10^{42}$~erg s$^{-1}$ for
which we were able to measure reliable \sigmastar\ values.  The BH
masses are derived indirectly from \sigmastar, assuming the \msigma\
relation of Tremaine et al. (2002), while the bolometric
luminosities are derived indirectly from \loiii.  We show the
Magellan sample ({\it red open circles}), the Gemini sample from
\citet[][; {\it blue open squares}]{liuetal2009}, and the SDSS
sample ({\it small filled circles}).  Double nuclei from the Gemini
data are indicated with a cross.  
({\it b}) The inferred distribution of Eddington ratios for the entire sample, 
with the Magellan data ({\it red open}), Gemini data ({\it blue hatched}) and 
best-fit Gaussian ({\it thin solid}) overplotted.  We
note that the uncertainties in this derivation are very large; 0.4 dex
for the bolometric correction and 0.5 dex in \mbh.  Given such large
uncertainties, the \lledd\ distribution is strikingly narrow.  The
central values of the Kollmeier \etal\ ({\it dashed}), 
Gavignaud \etal\ ({\it long dash}), and Shen \etal\
({\it dash-dot})  distributions are shown for comparison.
\label{masslumall}}
\end{figure*}

It is hard to interpret observations of narrow-line AGNs in the
absence of direct measurements of BH mass. The targets may be
$10^8$~\msun\ BHs radiating close to their Eddington luminosity or
they may be more massive systems in a low accretion state.  Their
importance to the accretion history of the Universe, the
interpretation of the observed properties of the host galaxies, and
comparisons with broad-line objects all depend critically on the
assumed mass scale.  The stellar velocity dispersions and inferred
range of BH masses thus constitute one of the main results of this
paper.

Using the \msigma\ relation of \citet{tremaineetal2002}, we convert
\sigmastar\ into BH mass and in Figure 6 we plot \mbh\ versus
\loiii\ for the sample.  Lines of constant Eddington ratio are
highlighted, where the bolometric correction, ${\rm log}~L_{\rm bol}
= 0.99 {\rm log}~L_{\rm [O {\tiny III}]} + 3.5$ is taken from 
\citet{liuetal2009}.  This correction is derived in two steps.  First,
the conversion between \loiii\ and \lf\ is calibrated using SDSS
broad-line AGNs \citep{reyesetal2008}, and then the relation between
\lf\ and bolometric luminosity is derived based on two recent
compilations of spectral energy distributions (SEDs),
\citet{marconietal2004} and \citet{richardsetal2006}.  In the latter
case, we have extrapolated the UV-optical slope downward beyond the
near-infrared minimum in order to estimate the full contribution of
the big blue bump, but exclude the reradiated mid-infrared emission.
The estimated scatter is 0.4 dex, accounting for the 0.2 dex
systematic difference between the two chosen bolometric corrections,
the 0.36 dex scatter between \loiii\ and continuum luminosity in
broad-line objects, and the 0.05 dex scatter in the single-band
bolometric correction from Marconi \etal\ \citep{liuetal2009}.

As seen in Figure 6, our narrow-line quasars appear to span $\sim 2$
dex in BH mass, luminosity, and Eddington ratio. Nominally, the
Eddington ratios cluster around the value log~\lledd$=-0.7$ with a
dispersion of 0.7 dex (see the best-fit Gaussian fit to the
distribution in Figure 6{\it b}), but it is important to bear in mind how
the Eddington ratios are obtained. As was mentioned above, there is an
0.4 dex uncertainty in the conversion from \loiii\ to
$L_{\rm bol}$. Furthermore, there is a $\sim 0.5$ dex uncertainty associated
with each \mbh\ measurement which arises from the uncertainty in the
velocity dispersion. Therefore, the observations are consistent with a
population of sources radiating close to their Eddington luminosities,
with the distribution of Eddington ratios broadened only by
observational uncertainty.

Now let us compare our observations with the masses and Eddington
ratios found for broad-line AGNs at similar redshifts.
\citet{kollmeieretal2006} find a strikingly narrow range in Eddington
ratio for luminous broad-line AGNs at $0.3 < z <
4$. \citet{shenetal2008} and \citet{gavignaudetal2008} report similar
results, although the latter paper finds evidence for an increased
spread in \lledd\ at lower luminosity.  It appears that our sample is
consistent with a similar behavior: rather than spanning a wide range
in accretion rate, the objects are clustered close to their Eddington
limits.  At face value, not only do the narrow- and broad-line objects
share comparable space-densities at the same luminosity
\citep{reyesetal2008}, but their distributions of mass and Eddington
ratio are also surprisingly similar.  Probably our application of a
luminosity threshold is partially responsible for the observed narrow
distribution in BH mass.  The Eddington limit effectively determines
the lowest-mass BHs to be included in the sample, while the space
density of high-mass BHs declines exponentially
\citep[e.g.,][]{hopkinsetal2009}.  Nevertheless, while a variety of
correlated errors or selection effects may contribute to the apparent
narrow distributions observed in each population, the point we
emphasize here is that the selections and biases are quite different
between the broad and narrow-line objects, and thus the similarity in
derived distributions is surprising.

There are several steps involved in calculating \mbh\ and \lledd, each
of which has the potential to introduce systematic biases into the
measurements. First, there is still substantial disagreement about the
calculation of \mbh\ in broad-line quasars on the basis of
continuum luminosity and broad line width
\citep[e.g.,][]{baskinlaor2005,collinetal2006,gavignaudetal2008,shenetal2008}.
Second, the bolometric luminosity correction for narrow-line quasars
uses a conversion from \loiii\ to $M_{\rm B}$ derived for broad-line
quasars, and this procedure may systematically bias the bolometric
luminosities towards low values \citep{netzeretal2006,reyesetal2008}.
Furthermore, one may question our assumption that the \msigma\
relation holds for narrow-line quasars. For one thing, they are active
(i.e., currently accreting) BHs and thus may be offset from the local
relation \citep{hoetal2008b,kimetal2008b}. In addition, there may or
may not be significant redshift evolution of the local \msigma\
relation that we use to calculate \mbh, and both the sign and the
magnitude of this change are at present unknown
\citep{wooetal2006,pengetal2006a,pengetal2006b,
  treuetal2007,laueretal2007}.  Each of these effects may potentially
bias the calculation of Eddington ratios by as much as 0.5 dex.  The
similarity of Eddington ratios of broad- and narrow-line quasars, as
well as the narrowness of their distribution, is an interesting and
suggestive result, but we caution that some systematic effects may be
affecting the comparison.

\section{Comparison of Stellar and Gaseous Kinematics}

It has long been known that the \oiii\ line width traces the bulge
velocity dispersion in low-luminosity AGNs
\citep[e.g.,][]{heckmanetal1981,whittle1992a,whittle1992b,nelsonwhittle1996}.
Recently, with the discovery of the \msigma\ relation
\citep{gebhardtetal2000a,ferraresemerritt2000}, interest in the gas 
velocity dispersion (\sigmagas)
has increased, due to the possibility of using it as
a proxy for \sigmastar\ in distant or luminous active galaxies for
which it is prohibitive to measure \sigmastar\ directly
\citep[e.g.,][]{nelson2000,wanglu2001,boroson2002,shieldsetal2003,grupemathur2004,onkenetal2004,greeneho2005o3,riceetal2006,salvianderetal2007}.
As a result, it becomes all the more crucial to understand the
physical origin of the observed correlation, and to characterize
whether it depends on properties of the AGN, such as bolometric
luminosity, Eddington ratio or radio luminosity.  This work increases
the luminosity range over which such comparisons have been done by
nearly two orders of magnitude, providing a crucial check of the
assumption that \sigmastar$\approx$\sigmagas.  By the same token,
given that \sigmagas\ and \sigmastar\ have now been measured for large
samples of weakly active galaxies over a wide range of Hubble types
\citep{ho2009o3}, it is possible to use the variations in this ratio
as a {\it probe} of the impact of the AGN on the surrounding interstellar 
medium (ISM).

\begin{figure*}
\vbox{ 
\vskip -7mm
\hskip 0.1in
\psfig{file=picture_jenny.epsi,width=0.8\textwidth,keepaspectratio=true,angle=0}
}
\figcaption[]{
({\it a}) Comparison of the velocity dispersion in warm gas as traced
by the [O {\tiny III}] line to that of the stars.  Following,
e.g,~\citet{nelsonwhittle1996} we plot the
FWHM/2.35 for the gas versus the stellar velocity dispersion, also
measured assuming a Gaussian broadening function (\S 3).  We
highlight the Magellan sample ({\it open circles}) and Gemini sample
of \citet[][; {\it open squares}]{liuetal2009}.  Error bars in 
\fwoiii\ are dominated by deviations from the assumption of 
Gaussianity, determined by normalizing the measured full-width at 
25\% and 75\% of maximum.
({\it b}) Same as ({\it a}) but using [O {\tiny II}] to measure the
gas line width.
({\it c}) Dependence of the deviation of \fwoiii\ from \sigmastar\ as
a function of \loiii.  Formally there is no correlation.  Symbols as
above, but in grey we have added the lower-luminosity sample of
\citet{greeneho2005o3} to increase the dynamic range.  Large filled
circles show the median and standard error calculated in 0.5 dex bins
for each subsample.  There is no clear correlation between
\delsig\ and $L_{\rm [O {\tiny III}]}$ even over this large
luminosity range.  ({\it d}) Same as ({\it c}), but using \fwoii.
\label{sigstarsiggas}}
\end{figure*}

As shown in Figure 7, there appears to be no correlation between
either \fwoiii\ or \fwoii\ and \sigmastar\ at these high luminosities
(the Spearman correlation coefficient is $\rho = 0.07$ with a
probability $P=0.4$ that no correlation is present).  We note that
while we span a narrow range in luminosity, we actually span a wide
range in \sigmastar, suggesting that the lack of correlation is a
generic property of luminous obscured AGNs, and may well be a generic
property of all AGNs at such high luminosities.  We try to salvage the
correlation by relying on lines with lower ionization potential, since
many studies have found systematic trends between line shape and the
critical density or ionization potential of the lines
\citep[e.g.,][]{pelatetal1981,filippenkohalpern1984,filippenko1985,
  derobertisosterbrock1986}.  Generically speaking the high-ionization
and/or high-critical density lines originate from closer to the BH and
are thus typically broader, making low-ionization lines a better
choice as proxies to \sigmastar\
\citep[e.g.,][]{greeneho2005o3,komossaxu2007}.  Instead, we find that
the \oii\ and \oiii\ gas widths are strongly correlated ($\rho = 0.9$,
$P=10^{-5}$; \fwoii/\fwoiii$=1.0 \pm 0.2$), but neither correlates
well with \sigmastar.

In search of some physical insight into the processes driving the gas
dispersion, we explore correlations between various AGN properties and
\sigmagas/\sigmastar.  For convenience, we define the logarithmic
difference \delsig$ \equiv $log \sigmagas$-$log \sigmastar.  Within
the sample as a whole, $\langle$\delsig$\rangle = 0.04 \pm 0.2$.  This
is to be compared with $\langle$\delsig$\rangle = -0.15 \pm 0.22$ for
low-luminosity AGNs in early-type galaxies \citep{ho2009o3} and
$\langle$\delsig$\rangle = 0 \pm 0.46$ for local narrow-line AGNs from
SDSS \citep{greeneho2005o3}.  Historically, many papers have found a
correlation between radio power and \oiii\ line width
\citep[e.g.,][]{wilsonwillis1980,whittle1992a,whittle1992b,nelsonwhittle1996}.
We see no evidence for increasing \delsig\ or \sigmagas\ with
increasing radio power over a range of $10^{22} < L_R$/\whz$ <
10^{25}$ at 1.4 GHz.  The Nelson \& Whittle sample does extend an
order of magnitude higher in radio luminosity, but it is clear that
jets alone cannot explain the sources with very large gas dispersions
\citep[see also][]{greeneho2005o3,ho2009o3}.  According to
\citet{ho2009o3}, the deviation of \sigmagas\ from \sigmastar\ shows a
clear dependence on AGN luminosity and Eddington ratio \citep[see
also][]{greeneho2005o3}, in the sense that more luminous AGNs have a
larger \sigmagas/\sigmastar\ ratio on average.  We see no
statistically significant correlation between \loiii\ and \delsig\
($\rho = -0.1$, $P=0.2$).  Granted, the dynamic range in luminosity is
very narrow for this sample alone, but, if anything, there is a weak
negative trend of \delsig\ with luminosity.  Indeed, the
lower-luminosity data from \citet{greeneho2005o3} shown in Figure
7{\it cd} clearly demonstrate that there is not a well-defined
correlation between \delsig\ and \loiii\ over a much larger luminosity
range.

While these results are discouraging in terms of the use of \fwoiii\
to approximate \sigmastar, they do tell us in no uncertain terms that
the ISM in the hosts of these luminous quasars is very disturbed.  As
mentioned above, gas line widths in the centers of inactive and weakly
active galaxies are well-described by virial or slightly sub-virial
motions due to turbulence \citep{ho2009o3}.  In low-luminosity active
galaxies, there is some evidence that AGN activity is, on average,
increasing the gas line widths \citep{greeneho2005o3,ho2009o3}, but
overall the gas still appears to be dominated by virial motion, at
least in a luminosity-weighted sense.  In stark contrast, at the
highest AGN luminosities, we see no correlation of gas and stellar
velocity dispersions, suggesting that the ISM is completely disturbed.

\vskip 3mm
\hskip -0.1in
\psfig{file=tablesfrv7.epsi,width=0.52\textwidth,keepaspectratio=true,angle=0}
\vskip 4mm

\section{Coincidence of Star Formation and Black Hole Growth}

Despite decades of research, the prevalence of temporally coincident 
star formation and accretion remains one of the outstanding problems in 
AGN physics \citep[e.g.,][]{borosonoke1984,heckmanetal1995,canalizostockton2001,
  kauffmannetal2003agn,ho2005b}.  Generically speaking, it seems
difficult to inhibit star formation when there is a large enough gas
supply to fuel a luminous AGN.  On the other hand, robust measurements
of star formation are difficult to obtain, particularly with a
luminous broad-line AGN spilling light in all directions and at all
wavelengths.  Our hope is to gain new insights from the study of 
obscured but intrinsically quite luminous accreting BHs.

Of course, there are always complications.  In our case, although the
primary emission from the AGN is obscured, nevertheless accretion
luminosity manages to contaminate many standard star formation
indicators.  The most reliable instantaneous (i.e. $<10^7$ yr) rate
diagnostic, \halpha\ luminosity \citep[e.g.,][]{kennicutt1998}, is
excited by accretion power rather than young stars in our case.
Likewise, blue/UV light is prone to contamination from scattered AGN
light, whose magnitude can be very significant at these high
luminosities
\citep[e.g.,][]{heckmanetal1995,zakamskaetal2005,zakamskaetal2006,youngetal2009,liuetal2009}.
In what follows we show that all of our galaxies are bluer than their
inactive counterparts, and we use a variety of indirect techniques to
disentangle the young starlight from scattered AGN light.

We begin by modeling the SDSS colors using stellar population
synthesis techniques.  As input, we use the SDSS $ugriz$ magnitudes,
with the \oiii\ and $k$-corrections applied, as described in \S 4.3.
We do not apply any correction for potential scattered light, as our
goal is to establish whether or not there is a blue component apparent
in the integrated colors of the galaxies.  The stellar population
synthesis code is described in detail in \citet{conroyetal2009}.  In
short, the code utilizes isochrones from Padova
\citep{marigogirardi2007}, the BaSeL3.1 stellar libraries with
thermally pulsating asymptotic branch stars added
\citep{lejeuneetal1997,lejeuneetal1998,westeraetal2002}, and (in this
case) a \citet{kroupa2001} initial mass function.  Dust attenuation
optical depths are allowed to vary separately for old and young stars,
following the model of \citet{charlotfall2000}.  

\psfig{file=d4000hdelta_sdss.epsi,width=0.45\textwidth,keepaspectratio=true,angle=0}
\vskip -0mm \figcaption[]{
The \dn\ and \hd\ indices measured from 
our continuum fits (see text for details).  The locus of points from 
\citet{kauffmannetal2003a} is shown in contours, while our measurements 
are the overplotted red crosses.  Stellar age increases towards the bottom 
right of the diagram.  The top histogram shows the 
distribution of \dn\ from the spectra themselves corrected for 
[Ne {\tiny III}], 
but not corrected for the blue power-law component while the points are 
derived from the spectral modeling.
\label{cfsigma}}
\vskip 5mm
\noindent
The fraction of total
stellar mass formed in the last 300 Myr for each galaxy is shown in
Table 4, with a median value of $1 \%$ at a median stellar mass of
$10^{11}$~\msun.  Already we see that while the BHs are experiencing a
major growth episode, the galaxies are not.

The integrated colors show that blue light is prevalent, but
spectroscopy is required to determine its origin.  For the purpose of
comparing with \citet{kauffmannetal2003a}, we measure the \dn\ and
\hd\ indices.  This two-dimensional diagnostic is predominantly
sensitive to age, and relatively insensitive to variations in
metallicity or dust.  We follow their conventions, wherein \dn\ is
defined as the ratio of fluxes in bands at 3850--3950 and 4000--4100
\AA\ \citep{baloghetal1999}, and \hd\ is defined in
\citet{wortheyottaviani1997}. As a function of age, galaxies follow a
well-defined sequence in \dn-\hd\ space, and the locus of SDSS
galaxies in this space is shown in Figure 8.  Unfortunately, because
our AGN-powered emission lines have such high EWs, we cannot measure a
meaningful \hd\ index directly from the spectra.  We can measure \dn\
after correcting for [Ne {\small III}]$~\lambda 3869$, and those
values are shown in the top histogram in Figure 8.  [We have verified
that we can reproduce the values in Kauffmann et al.  Measuring the 30
most luminous objects in their DR2 catalog, we find
$\langle$[\dn$_{\rm our}$-\dn$_{\rm Kauff}$]/\dn$_{\rm Kauff} \rangle=
-0.04 \pm 0.05$.]  The median of our sample, uncorrected for any blue
component, is $\langle$\dn$\rangle = 1.2$.  For comparison, the thirty
most luminous targets in the Kauffmann \etal\ DR2 sample, with a
median \loiii\ of $10^{42}$~erg~s$^{-1}$ ($\sim 0.5$ dex less luminous
than our median luminosity), have $\langle$\dn$\rangle = 1.3$.

Our first task is to measure the luminosity of the blue continuum.
The LDSS3 spectra do not always include the 4000~\AA\ break, and so we
use the SDSS spectra for this measurement.  We wish to model the
various spectral components, but to mitigate degeneracy we restrict
the model to nothing more than a K star, an A star, and a blue power
law whose slope is fixed to $P_{\lambda} \propto \lambda^{-1}$.  In
principle the slope of the blue component will depend on the physical
process responsible for it; given the modeling degeneracies and the
uncertainties in internal reddening and flux calibration, we do not
attempt to measure the true spectral shape of the blue light. Likewise
\sigmastar\ is fixed to the best-fit value determined in \S 3.  While
the fits are not unique, we believe each component is reasonably well
constrained: a significant A or F star contribution is fixed by
higher-order Balmer absorption lines, while the EW of the old stellar
component provides a direct probe of the dilution by blue light.  By
repeating the fitting with a power-law slope of $-1.5$, we confirm
that the derived luminosities are insensitive to the assumed power-law
slope.  More reassuringly, fits performed without an A star component
yield blue luminosities that agree within 0.05 dex of the nominal
result.

\begin{figure*}
\vbox{ 
\vskip -2mm
\hskip 0.5in
\psfig{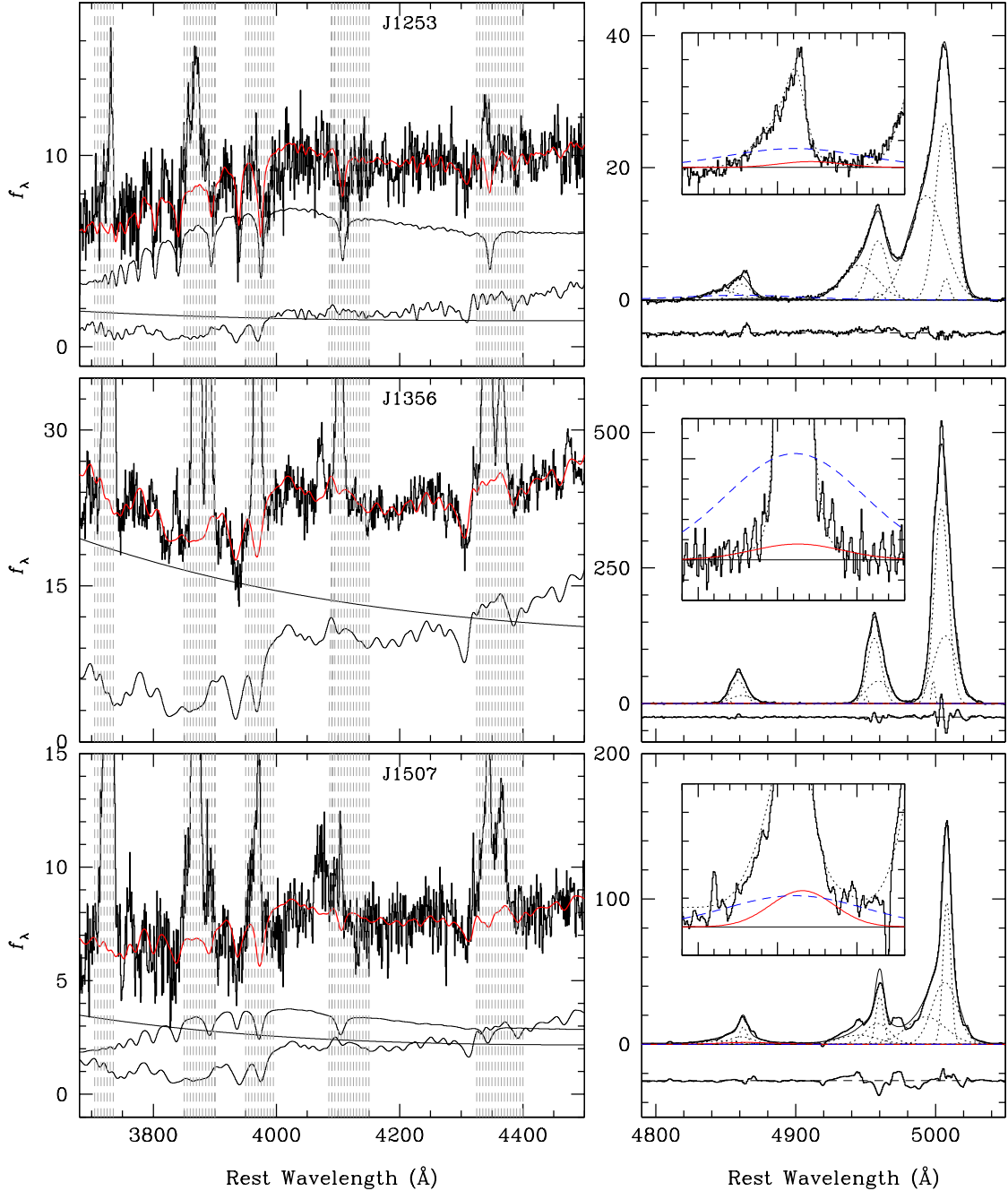}
}
\vskip -0mm \figcaption[]{ 
({\it Left}) Three example fits used to constrain the blue-light fraction.
We fit $3680-5400$ \AA\ with a model including a K star, an A star, and 
a power law ($P_\lambda \propto \lambda ^ {-1.5}$), while \sigmastar\ is 
fixed to our best fit from above.  We plot the SDSS data 
({\it solid histogram}), the best-fit model ({\it red line}), 
and model components ({\it thin black solid line}) as well as excluded regions 
({\it grey hatched}).  J1253 and J1507 contain a mixture of all three 
components, while J1356 requires only a K star and a (significant) 
power-law component.
({\it Right}) We show our best fits to the [O {\tiny III}]+\hbeta\
region of our LDSS3 spectra, with narrow \hbeta\ constrained to
have the same shape as [O {\tiny III}].  We show the data ({\it
solid histogram}), the total model ({\it thin solid line}), each model
component ({\it dash-dot line}), the broad \hbeta\ component ({\it red
solid line}), and the expected scattered broad component ({\it blue
dashed line}).  The latter assumes that all of the blue light is scattered, 
takes a standard conversion between \hbeta\ and \lf\
\citep{greeneho2005cal}, and assumes an \hbeta\ linewidth of 
4000 \kms.  In J1356 and J1253 there is evidence for
recent star formation from the excess of blue light compared to the
apparent broad component at \hbeta, while J1507 can be explained
purely with scattered light.  Flux density in units of
$10^{-17}$~\flamb.
\label{cfsigma}}
\end{figure*}

These fits provide two useful measurements: one is a blue luminosity,
and the other is a (model-dependent) measurement of the \dn\ and \hd\
indices for the underlying stellar population, as shown by red crosses
in Figure 8.  The stellar population gets older as one moves to the
lower right of this diagram.  Note that three objects have \dn\
indices that are larger than values seen in real galaxies.  Presumably
the discrepancy arises because in these cases the stellar continuum is
modeled by a single K star, whereas in real stellar populations there
will be some contribution from M stars as well as younger populations.
The \dn\ and \hd\ indices derived this way represent a lower
limit on the young population, while below we will attempt to evaluate
what fraction of the blue light may originate in young stars.

If the blue light is predominantly scattered AGN light, we can predict
the corresponding reflected broad \hbeta\ component, using the
well-known tight correlation between optical/UV continuum luminosity
and Balmer emission \citep[e.g.,][]{searlesargent1968}; we use the
recent calibration of \citet{greeneho2005cal}.  The expected line
fluxes are low ($\sim 10^{-16}$~erg~s$^{-1}$~cm$^{-2}$), but we can at
least determine whether or not our spectra are consistent with a broad
component at this level.  In the most luminous narrow-line quasars,
\citet{liuetal2009} robustly detect a broad \hbeta\
component in a significant fraction of their sample and determine the
scattered light contribution to the blue continuum \citep[see
also][]{terlevichetal1990,heckmanetal1995,cidfernandesetal2004}.
Here, because the lines are much weaker, we rather take the indirect
approach of comparing the expected scattered emission from the
continuum fits with a direct fit to the \hbeta\ line.  If the two
\hbeta\ fluxes are consistent, then we conclude that the blue light
may be ascribed exclusively to scattered light.

Since the narrow-line shapes are quite complex and highly
non-Gaussian, care must be taken to ensure that structure in the
narrow lines does not masquerade as a bona fide scattered broad
component.  We thus repeat the narrow-line fits as described above,
except that in this case we fix the narrow components of \hbeta\ to
that of \oiii, and allow the fit to include an additional broad
\hbeta\ component.  We constrain $3000<$FWHM$_{\rm \hbeta}<6000$~\kms\
(Liu et al. find \hbeta$_{\rm FWHM}\approx 4000$~\kms), since
otherwise the broad line will often attempt to fit low-level wiggles
in the continuum.  

\begin{figure*}
\vbox{ 
\vskip -0.1truein
\hskip 0.in
\psfig{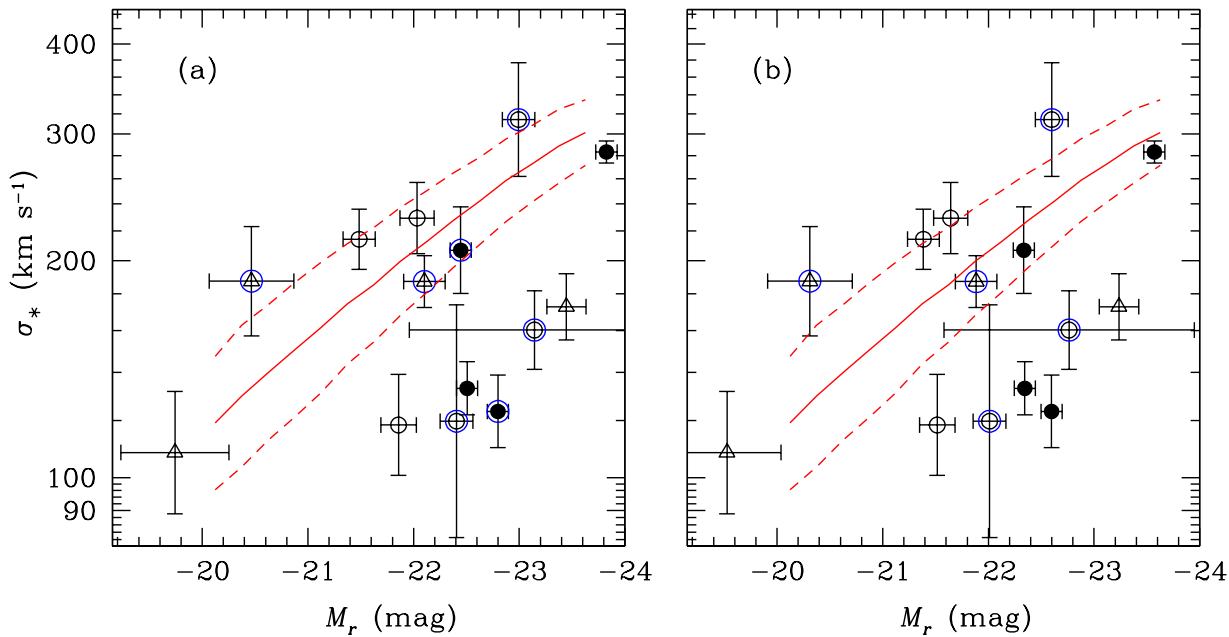}
}
\vskip -0mm \figcaption[]{ 
({\it a})Rest-frame $r$-band Faber-Jackson relation.  
We differentiate single $n>2$
profiles ({\it open circles}), bulge components of disk galaxies ({\it
triangles}), and disturbed systems for which we use the nonparametric
Petrosian magnitude ({\it filled circles}).  Active galaxies with
\loiii$>10^{42.5}$ erg~s$^{-1}$ are circled in blue.  The inactive
relation ({\it red solid line}), and uncertainties thereof ({\it red
dashed lines}), are from \citet{desrochesetal2007}, and are shown only
over the region for which they were measured.  
({\it b}) The same as ({\it a}) but with both a correction for the blue light
(see text) and an evolution correction of 0.9$z$ mag.
\label{fj}}
\end{figure*}

In all cases the \oiii\ and \hbeta\ line shapes
match each other extremely well (Fig. 9), and in eight cases we detect
a broad component (Table 4; Fig. 9 {\it bottom}).  We convert the rms
residuals into an upper limit on broad \hbeta\ assuming a FWHM of 4000
km s$^{-1}$.  We take as the upper limit on the scattered broad
component either the detected broad \hbeta\ luminosity or the upper
limit thereof, whichever is larger, and then compare with the
luminosity we expect based on the detected blue continuum.  If the two
agree within a factor of 2, then we believe scattered light can fully
account for the blue light, whereas when the continuum level is
higher, we ascribe the excess to current star formation (the five
cases are noted with * in Table 4).

When detected, the blue light accounts for anywhere from 5\% to 60\%
of the total continuum flux at 4000 \AA.  After the blue light is
removed, our galaxies occupy a sensible region of the \hd-\dn\ plane
relative to the inactive galaxy population. The median \dn\ rises from
1.2 to 1.6. Even excluding those galaxies with \dn\ larger than what
is seen in SDSS galaxies yields an average value of 1.5.  In other
words, while there is clearly an age spread, the corrected index
values suggest that the galaxy light is dominated by old stellar
populations in the majority of cases.  In five galaxies (J0841, J1222,
J1253, J1356, and J2212), we find a blue continuum in excess of what
can be explained by scattered light, which we interpret as a star-forming 
component.  It is interesting to note that these systems tend
to have higher star formation rates inferred from their broad-band
colors as well.  Even more intriguing is the observation that all of
the highly disturbed systems have an excess (star-forming) blue
component.  On the other hand, we see no correlation between the \dn\
values and host morphology. In summary, these host galaxies are
forming $\lesssim 1 \%$ of their stellar mass at the present time.
About half of the objects are consistent with having only old stellar
populations, while the others show evidence for ongoing (in the form
of excess blue light) or recent [low values of \dn] star formation at
a low level.  While we cannot place absolute limits on the current
star formation rate to any great precision, we argue here, and based
on the structural parameters investigated below, that the galaxies are
effectively fully formed.

In Figure 10 we show the $r$-band Faber-Jackson relation for the
galaxy bulge components, with the most luminous targets circled in
blue and the disk- or bulge-dominated galaxies differentiated as
triangles or circles, respectively.  Obviously there is a large amount
of scatter in this diagram, much of it measurement-based.  It appears
that nearly half of the targets are offset toward more luminous bulges
at a given \sigmastar, but we caution that there are substantial
systematic biases hidden in the bulge luminosities, as discussed
below.

We do not see any correlations between location in the
Faber-Jackson diagram and AGN luminosity, young stars (as traced by
the blue light), or intermediate-age stars [\dn].  Partly, the sample
is just too small, spanning too narrow a range in luminosity, to
display any such trends.  At the same time, a number of potential
uncertainties may impact the measured luminosities.  There is
firstly the contamination from scattered blue light, which boosts the
observed $r$-band magnitudes by as much as 0.5 mag.  Also, the targets
span a range of redshifts ($0.1 < z < 0.45$), but we know that over
time a collection of stars will become less luminous as the most
massive stars evolve.  Unfortunately, as shown by, for example,
\citet{conroyetal2009}, the precise magnitude of this fading is highly
uncertain (by $\sim 1$ magnitude), due to our inadequate knowledge of
the slope of the initial mass function around the turnoff mass. Taking
a value of 0.9$z$ mag in $r$ for the purposes of illustration
\citep[e.g.,~][]{vandokkumfranx2001,treuetal2005,conroyetal2009}, and
removing the detected blue light yields the magenta points,
substantially lowering the discrepancy with the local Faber-Jackson
relation. We are reluctant to provide any additional interpretation,
given the small sample and substantial uncertainties in the
measurements.

Our basic conclusion, combining the results of the stellar population
modeling, stellar population indices, and structural properties, is
that for the most part the host galaxies had their stellar content in
place prior to this epoch of substantial BH growth.  The general
result, that we see a range of star-forming properties with no clear
correlation between host galaxy morphology and star formation
properties, is in broad agreement with previous results for
lower-luminosity narrow-line AGNs
\citep[e.g.,][]{schmittetal1999,gonzalezdelgadoetal2001,cidfernandesetal2004}.

Most interestingly, the situation appears rather different for objects
at the highest luminosities.  There, \citet{liuetal2009}
see clear evidence for
Wolf-Rayet features in $\sim 1/3$ of their targets (we see none in
this sample), which constitute a clear signature of a recent burst of
star formation \citep[e.g.,][]{allenetal1976,kunthsargent1981}.
Furthermore, \citet{zakamskaetal2008} find very high star formation
rates from mid-infrared spectroscopy, for a different sample of
luminous narrow-line AGNs, rates that are substantially higher than
either those seen in inactive galaxies of similar mass or those in
typical broad-line AGN samples of similar luminosity.  Now,
\citet{kauffmannetal2003agn} find compelling evidence that the 
post-starburst fraction increases with increasing \oiii\ luminosity.
We may be seeing a hint that the galaxy growth rate scales with AGN
luminosity even up to starbursts at the highest AGN luminosities
\citep[see also][]{
canalizostockton2001,vandenberketal2006,letaweetal2007,jahnkeetal2007}.

\hst\ images show that hosts of luminous narrow-line quasars from the
SDSS-selected sample are bulge-dominated \citep{zakamskaetal2005},
whereas those of somewhat less luminous (by about 0.5 dex) IR-selected
narrow-line quasars are dusty, disky systems
\citep{lacyetal2007}. This paper presents observations of objects with
luminosities very similar to those in our HST sample, with similar
results: about a third of the objects are found in strongly disturbed
interacting systems, while the rest are found in bulge-dominated
galaxies. We previously suggested that the difference in morphology
may reflect the difference in the quasar fueling mechanisms as a
function of quasar luminosity \citep{zakamskaetal2008}, but of course
this interpretation is only tentative since the comparison samples were
selected using very different methods. A detailed study of hosts of
optically selected narrow-line quasars at lower luminosity is needed
to understand the origin of the morphological difference.

\section{Discussion and Summary}

We have chosen to explore the host galaxy properties of obscured
active galaxies in particular because of the increased contrast
between galaxy and nucleus.  On the other hand, it is our burden to
determine whether or not our conclusions can be generalized to all
accreting BHs at the epoch being probed.  Rather surprisingly, the
distributions of BH mass and Eddington ratio for our sample of
narrow-line AGNs are quite similar to those seen in broad-line samples
at similar luminosity and redshift
\citep[e.g.,][]{kollmeieretal2006,shenetal2008,gavignaudetal2008}.
This fact, combined with the observation of \citet{reyesetal2008} that
the space densities of the broad- and narrow-line objects are
comparable at fixed luminosity, suggest that the two may originate
from similar populations.  There is another aspect to the story, as
illuminated by the differences in star-forming properties between the
narrow- and broad-line objects.

A variety of recent papers have devised clever techniques for
disentangling photoionization by stellar and nonstellar sources using
the line strengths of high- and low-ionization lines, where the former
are certainly dominated by the AGN but the latter can be significantly
contaminated by star formation
\citep[e.g.,][]{ho2005a,kimetal2006,villarmartinetal2008,
  melendezetal2008}.  These studies all come to the same conclusion;
star formation is more energetically significant in narrow-line than
in broad-line AGNs, particularly at the luminous end \citep[see
also][]{zakamskaetal2008}.  The excess star formation in narrow-line
objects has been interpreted as evidence for an evolutionary scenario,
in which obscured accretion and ongoing star formation coexist, and
then accretion terminates further star formation, yielding an
unobscured object
\citep[e.g.,][]{sandersmirabel1996,canalizostockton2001,
  ho2005a,hopkinsetal2006}.  In this picture some event (usually a
merger) induces both the star formation and nuclear activity, and the
broad-line object signifies the final stage in the accretion cycle.

Perhaps, on the other hand, there is not necessarily a causal
connection between the star formation event and the accretion event,
but when both occur simultaneously the odds are higher that the
accretion event will be obscured by star formation in the host galaxy
\citep{rigbyetal2006,lacyetal2007}. Our data alone clearly cannot
distinguish between these two possibilities.  However, we do
see a range of star formation properties across even this relatively
small sample.  Naively, if there were truly a causal connection
between star formation and nuclear activity, one would expect that the
majority of our sample would show a clear sign of ongoing star
formation, even if only in the form of post-starburst signatures.
We look forward to analyzing the full \citet{reyesetal2008}
sample to more definitively quantify the star formation properties as
a function of AGN luminosity \citep{liuetal2009}.

We have seen that luminous obscured active galaxies with $0.1 < z <
0.45$ and $10^{43} < L_{\rm bol} < 10^{46.5}$~erg~s$^{-1}$ are
composed of BHs with masses $\sim 10^8$~\msun\ radiating at a high
fraction of their Eddington limits.  We are seeing unambiguous
evidence that BH accretion impacts the host on a galaxy-wide scale;
the host galaxy ISM is clearly substantially disturbed relative to
less active systems.  As a result, we cannot recommend the use of gas
line widths as a substitute for stellar velocity dispersions at such
high luminosities, at least not in obscured sources.  There is a clear
blue component in the galaxies, apparent both in the overall colors
and in nuclear spectroscopy, and in most cases the blue light is
consistent with being scattered emission from the obscured nucleus.
Nevertheless, nearly half of the sample shows evidence for recent or
ongoing star formation, while the rest are consistent with
predominantly old stellar populations.  Although star formation
activity may rise with vigorous AGN activity, there is a wide
dispersion at all luminosities, suggesting that we are far from having
a consistent picture of the coevolution of host galaxies and BH
growth.

\acknowledgements{  
We are grateful to D.~Kelson and G.~Walth for significant assistance
using COSMOS, while T.~Matheson kindly provided the IDL routines
used to reduce the spectra. We thank M. Strauss for many useful
conversations, C. Conroy for operating his new stellar population
code for us, and R.~Reyes for assistance deriving bolometric
corrections.  The referee provided useful suggestions that improved
the quality of this manuscript.  We acknowledge the exquisite
support of the staff at Las Campanas and, finally, we thank the SDSS
team for providing the incredibly rich data set upon which this work
is based. Support for J.~E.~G. was provided by NASA through Hubble
Fellowship grant HF-01196 awarded by the Space Telescope Science
Institute, which is operated by the Association of Universities for
Research in Astronomy, Inc., for NASA, under contract NAS 5-26555.
Research by A.~J.~B. is supported by NSF grant AST-0548198. }

\end{document}